\documentclass[a4paper,12pt]{article}

\usepackage[margin=2.65cm]{geometry}
\usepackage[utf8]{inputenc}
\usepackage{amsmath}
\usepackage{natbib}
\usepackage{graphicx}
\usepackage{url}
\usepackage{rotating}

\begin{document}
\title{Correcting for non-ignorable missingness in smoking trends}
\date{20th November 2014}
\maketitle
\begin{center}

\author{Juho Kopra  \\
	Department of Mathematics and Statistics,\\
	University of Jyväskylä, Finland  \\[0,5cm]
	\and 
	Tommi Härkänen \\
	National Institute for Health and Welfare,\\
	Helsinki, Finland \\[0,5cm]
	\and 
	Hanna Tolonen \\
	National Institute for Health and Welfare,\\
	Helsinki, Finland \\[0,5cm]
	Juha Karvanen \\
	Department of Mathematics and Statistics,\\
	University of Jyväskylä, Finland
	}


\end{center}

\setcounter{page}{1}


\setcounter{tocdepth}{2}

\section*{Abstract}
Data missing not at random (MNAR) is a major challenge in survey sampling. We propose an approach based on registry data to deal with non-ignorable missingness in health examination surveys. The approach relies on follow-up data available from administrative registers several years after the survey. 
For illustration we use data on smoking prevalence in Finnish National FINRISK study conducted in 1972-1997. The data consist of measured survey information including missingness indicators, register-based background information and register-based time-to-disease survival data. The parameters of missingness mechanism are estimable with these data although the original survey data are MNAR. The underlying data generation process is modelled by a Bayesian model. 
The results indicate that the estimated smoking prevalence rates in Finland may be significantly affected by missing data.

\section{Introduction} 
Participation rates in health examination surveys (HES) have been declining over the years in many countries. The declining participation rates inflict the estimation of health indicators in many ways. First, the low participation rates compromise the population representativeness of the sample because the participants and non-participants differ from each other. 
The non-participants are more often smokers \citep{shahar-1996,tolonen-2005} and have higher risk of death \citep{jousilahti-2005,harald-2007} compared to the participants. 
It has also been found that the non-participants tend to be men \citep{vanloon-2003,sogaard-2004}, younger persons \citep{sogaard-2004} and single \citep{shahar-1996,sogaard-2004,tolonen-2005}. Generally, the non-participants have been found to have lower socio-economic status \citep{jackson-1996,vanloon-2003,drivsholm-2006,harald-2007} and lower education  \citep{shahar-1996,sogaard-2004,tolonen-2005} than the participants. 
Second, the declining trends in participation rates may distort the trends of the estimated health indicators. Especially, if smokers, heavy alcohol users and obese are less eager to participate than they were decades ago, the trends of the health indicators may look more positive than they should.

In statistical terms, data from HES are missing not at random (MNAR) and consequently the missingness mechanism cannot be ignored in the analysis \citep{little-rubin-2002}. Although dealing with non-ignorable missingness is challenging in general, there are some methods for this. 
One of these is making functional assumptions for the joint distribution of missing data and observed values \citep{little1993pattern,ekholm1998muscatine}. 
This is usually accompanied with a sensitivity analysis for evaluating the effect of assumed missingness mechanism \citep{vanbuuren1999multiple}. If study design is longitudinal, the modelling of non-ignorable missingness may be based on partially available repeated measurements \citep{ibrahim2009missing}. 
Recently, a subsample ignorable likelihood (SIL) approach \citep{little-zhang-2011} was proposed for situations, where full data are available for some variables while the other variables have missing data.

We propose an approach to correct for non-ignorable missingness in situations where follow-up data are available for both participants and non-participants. 
Finland is one of the few countries where follow-up data for the entire survey sample can be obtained through a record linkage to the administrative registers. 
Naturally, the follow-up data will not be available right after the survey but only many years later. Without further assumptions, the trends of health indicators can be therefore corrected only retrospectively.

As an illustration for our approach, we use the data from the National FINRISK studies \citep{laatikainen-2003,harald-2007}, which are one of the data sources used to evaluate public health in Finland. 
The data from the surveys carried out in 1972, 1977, 1982, 1987, 1992 and 1997 are included. The participation to the physical measurements have decreased from 95\% in 1972 to 74\% in 1997. Note that in the next section we define participation differently. Under the decreasing participation, we estimate the prevalence of smoking utilizing the follow-up data available from the registers. 

The relevant details of the FINRISK surveys are presented in Section \ref{data}. In Section \ref{model}, a Bayesian model is built for the analysis of non-ignorable missing data. Section \ref{comparison-of-trends} compares the trends for non-ignorable and ignorable approaches, and Section \ref{discussion} concludes the paper.

\vskip14pt
\section{FINRISK data and linked register data} 
\label{data}
The National FINRISK Study (earlier North Karelia Project) data arose from a setup where the original aim was to intervene to people of North Karelia via a health education campaign. 
Later, the data have been collected every five years to measure the risk factors of key diseases and to monitor public health. In addition to North Karelia, the neighbour province of Northern Savonia has been included in studies since the beginning. Later, Turku and Loimaa area, Helsinki and Vantaa area and Oulu province have joined the survey. The data from the surveys conducted in  1972-1997 are used in this paper.

Sampling frame for the surveys has been the National Population Register\nocite{}. The survey design has changed over the study years, see Table \ref{sampling-design-cohorts}, but at each study the sampling has been stratified among the participating areas. In 1972, the sampling was systematic on birthdays, and people aged between 25-59 were sampled. In 1977 the simple random samples was drawn from people aged between 30-64. In 1982, the survey was balanced between the 10-year age-groups and 25-64-years-old people were sampled. In years 1987, 1992, and 1997 the sampling design was balanced sampling between 10-year age-groups within genders. In 1997, the eligible age was extended to 25-74-years-old in North Karelia and in Helsinki and Vantaa area.

\setlength{\tabcolsep}{5pt}
\begin{sidewaystable}[!htbp] \centering 
  \caption{Description of sampling design and eligible cohorts and areas. Area codes are; North Karelia: 2, Northern Savonia: 3, Turku and Loimaa: 4, Helsinki and Vantaa: 5, Oulu province: 6 (* = Upper eligible age is 75 for areas 2 and 5.). Marginal participation percentages are given for the studies. Two rightmost columns indicate the survey sample size and the corresponding count of observed lung cancer and COPD events for persons selected to the sample over the follow-up period.}
  \label{sampling-design-cohorts} 
\begin{tabular}{cllp{7cm}crr}
\\[-1.8ex]\hline 
\hline \\[-1.8ex] 
Year & Cohort & Areas & Sampling design & Participation & Sample size & Events\\\hline \\[-1.8ex]
1972 & 25-59 & 2,3 & Systematic on birthdate, balanced between areas & 86.0\% & 12377 & 420\\
1977 & 30-64 & 2,3 & Simple random sampling, balanced between areas & 88.1\% & 11319 & 373\\
1982 & 25-64 & 2,3,4 & Balanced between 10-years age-groups within areas & 80.0\% & 11332 & 281\\
1987 & 25-64 & 2,3,4 & Balanced between 10-years age-groups within areas within gender & 79.9\% & 7893 & 127\\
1992 & 25-64 & 2,3,4,5 & Balanced between 10-years age-groups within gender and areas & 76.2\% & 7895 & 97\\
1997 & 25-64* & 2,3,4,5,6 & Balanced between 10-years age-groups within gender and areas & 71.3\% & 11423 & 140\\\hline \\[-1.8ex]
\end{tabular}
\end{sidewaystable}

\begin{figure}[!htbp]
\includegraphics[width=\textwidth]{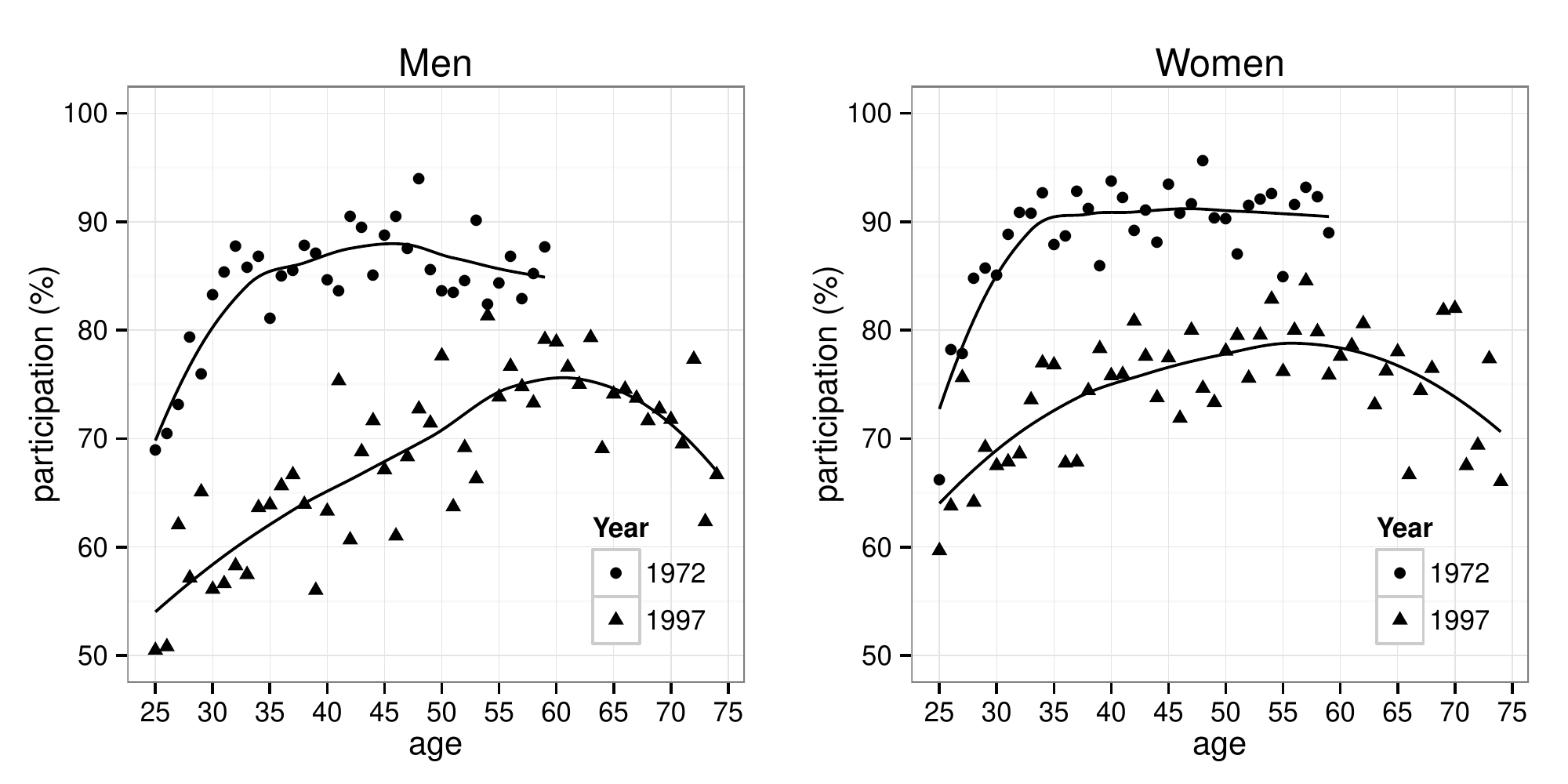}
\caption{Participation rate as a function of age in 1972 and 1997. Each circle and triangle represents the observed proportion of participants within one-year-age-group over all study areas studied that year. The graph shows that the participation rates have decreased in all age groups for both men and women. The solid lines are calculated using Locally Weighted Regression (LOESS) \citep{cleveland1979robust}.}
\label{data-osal}
\end{figure}

The participation is defined as answering to the question about daily smoking. This definition leads to lower participation rates than reported elsewhere because some individuals participated otherwise but skipped the smoking questions. The participation seem to depend on age and gender, but possibly also on smoking, which is to be investigated. 
The age-dependency of the participation rate and its change over the period 1972-1997 is shown in Figure \ref{data-osal}. 
Smoking, together with other health indicators, was measured by using a multi-page questionnaire. Smoking questions classified each person either non-smokers, ex-smoker or current smoker. We model smoking using two classes, where the ex-smokers and non-smokers are considered as the same.

\begin{figure}[!htbp]
\includegraphics[width=\textwidth]{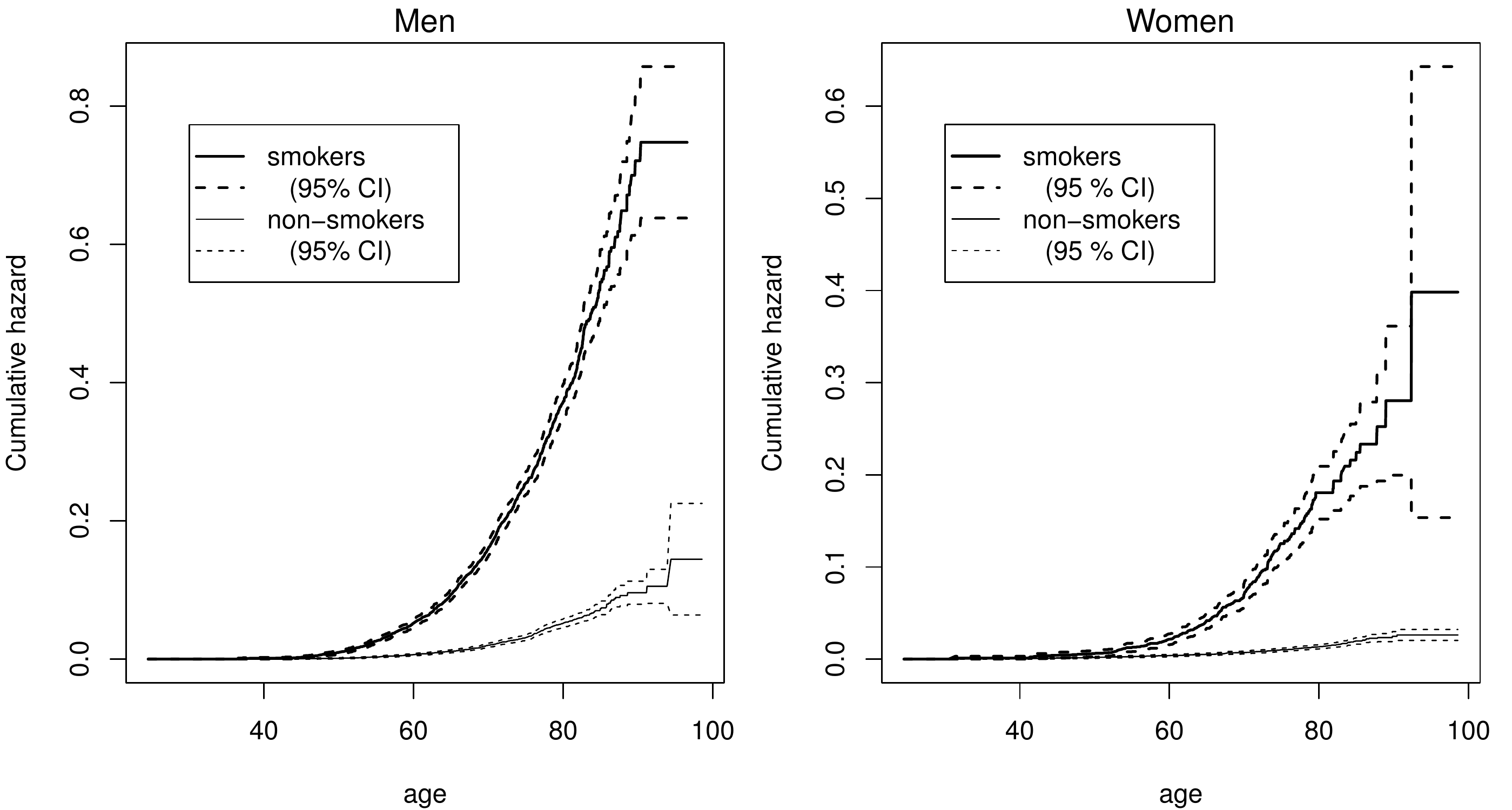}
\caption{Cumulative hazard estimates with confidence intervals for smoking-based diseases of lung cancer and COPD. Graphs are produced using the participant data only.}
\label{data-survival}
\end{figure}

The sources of the follow-up data are Care Register for Health Care (HILMO) \citep{careregister} and the cause of death data \citep{causeofdeath}. The follow-up data are linked to the survey data by personal identification number. 
The follow-up data contain the date and the cause of death, and the cause of hospitalization. 
The diseases considered here are lung cancer (ICD10: C34, ICD9/ICD8: 162) and chronic obstructive pulmonary disease (COPD) (ICD10: J41-J44, ICD9/ICD8: 491-492) for which smoking is known to be the main risk factor \citep{doll1956lung,wynder1994smoking,mannino2007global,cornfield2009smoking}. 
The follow-up data are available for all persons (participants and non-participants) selected to the FINRISK samples. 
The effect of smoking to the onset of lung cancer and COPD for men and women is illustrated in Figure \ref{data-survival}. 

We denote our variables as follows. 
The smoking indicator variable is denoted as $Y_i$ for person $i$. 
Background variables $X_i = (x_{1i},x_{2i},x_{3i},x_{4i})$ for person $i$ include the age at the beginning of the follow-up $x_{1i}$, area $x_{2i}$ and gender $x_{3i}$, which origin from the registers. The variable $x_{4i}$ is the study year. 

The sample indicator $m_{1i}=1$ indicates that person $i$ has been chosen to a survey sample, and participation indicator $M_{2i}=1$ indicates that he or she has participated to the survey. If $m_{1i}=0$ then $M_{2i}$ must also be $0$ because people outside of the survey sample can not take part. Variables $X_i$ are observed for both the participants and non-participants while $Y_i$ is observed only from the participants.

The follow-up data consist of time-to-event-variable $T_i$ and event indicator $r_i$, where $T_i$ is the age at the diagnosis of the disease, i.e. the onset of lung cancer or COPD. Variable $T_i$ is observed for the participants and non-participants. If a person has not been diagnosed until the end of follow-up period (31st December 2011), or if person dies for other causes, then the time-to-event-variable becomes right censored. 
In the case of right censoring we know only that $T_i > c_i$ where $c_i$ is persons age at censoring or age at death. The date of diagnosis can be the same as date of death, if person has not been diagnosed earlier and lung cancer or COPD is the cause of death. If person recovers from lung disease and becomes repeatedly diagnosed, the time-to-event-variable holds the time of the earliest diagnosis.

\section{Bayesian model for non-participation and smoking} 
\label{model}

\subsection{Dependency structure and modelling assumptions}
\begin{figure}[!htbp]
\includegraphics[width=\textwidth]{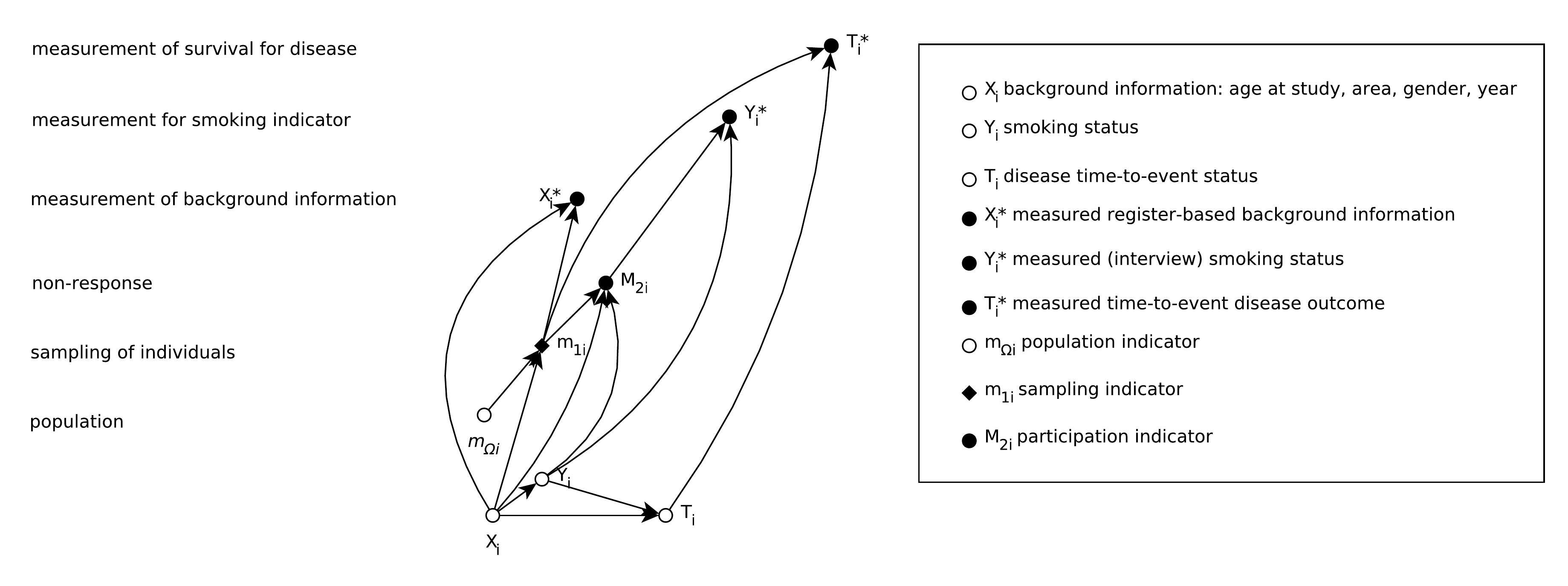}
\caption{Illustration of variable dependencies and the data-collection process.}
\label{model-structure}
\end{figure}

We present the structure of the model in Figure \ref{model-structure} using the concept of causal model with design \citep{karvanen-graphs-2014}. The Figure \ref{model-structure} represents a causal model at the bottom where background variables $X_i = (x_{1i}, x_{2i}, x_{3i}, x_{4i})$ affect the probability of smoking $P(Y_i)$ and the risk of lung disease $P(T_i)$. In addition, smoking also has an effect on the risk of lung disease. These relations are described as arrows $X_i \rightarrow T_i, X_i \rightarrow Y_i$ and $Y_i \rightarrow T_i$ in the model graph.
The causal relations of smoking and lung cancer \citep{doll1956lung,wynder1994smoking,cornfield2009smoking} and smoking and COPD \citep{mannino2007global} are known to exist. Also, it has been observed that the prevalence of smoking varies depending on the area, gender and age \citep{finriski2007_taulukkoliite,finriski2012_taulukkoliite}. 
Persons belonging to the sample have $m_{1i}=1$, and are selected from population $\Omega$, which in this case is the general Finnish population in geographically defined areas and age groups specified above. Sampling is based on the background register data, which is why  we have $X_i \rightarrow m_{1i}$ in Figure \ref{model-structure}.  Participation, which is indicated by $M_{2i}=1$, is affected by background variables ($X_i \rightarrow M_{2i}$) and smoking ($Y_i \rightarrow M_{2i}$). People may participate only if they are selected to the sample, which is indicated by the arrow $m_{1i} \rightarrow M_{2i}$ in the graph. If a person participates, he or she has $M_{2i}=1$, and thus $Y_i^* = Y_i$. Otherwise, smoking indicator is missing $Y_i^* = \text{NA}$. The background information as well as survival information $T_i^*$ are collected for all persons in the sample. The follow-up variable $T_i$ is a vector of two elements, the actual time variable $t_i$, either for the event time or censoring time, and an indicator variable for censoring, denoted as $r_i$. The notation for this is
\begin{equation}
T_i = (t_i,r_i) = \begin{cases} (t_i,0), & \mbox{if an event is observed} \\ (t_i,1), & \mbox{if an event is right censored.} \end{cases} \nonumber
\end{equation}
The observed $T_i^*$ is then defined as 
\begin{equation}
T_i^* = \begin{cases} T_i & \mbox{if person i belongs to a sample: } m_{1i}=1\\
\text{NA}, & \mbox{if person i does not belong to a sample: } m_{1i}=0. \end{cases} \nonumber
\end{equation}

The censoring due to deaths other than lung cancer or COPD is informative because smoking is a risk factor for many common causes of death. The usual way to deal with this kind of informative censoring is to define an additional endpoint for other deaths and use a competing risk  model \citep{kalbfleisch2002}. However, this would create new problems because we would implicitly assume that all differences in the mortality between participants and non-participants are due to smoking. In reality, participants and non-participants differ also by many other risk factors which work as confounders. Therefore, we have chosen to use only smoking-specific survival outcome in the analysis and to treat the censoring as non-informative. The implications to the results and alternative approaches are discussed in Section~\ref{discussion}.

In Figure \ref{model-structure}, the non-participation depends on smoking status $Y_i$, which means that the missingness mechanism is non-ignorable. In general, the non-ignorable missingness mechanism is not estimable from data. To overcome this issue, we use the follow-up data to make an additional assumption on the missingness mechanism.

We want to estimate the smoking prevalence for the whole sample, so we need to estimate the distributions 
\begin{equation}
P(Y_i) = P(M_{2i}=1)P(Y_i|M_{2i}=1)+P(M_{2i}=0)P(Y_i|M_{2i}=0) \hspace{20pt} i \in \Omega. \label{marginal_distY}
\end{equation} 
On the right hand side of Equation \eqref{marginal_distY} the probability of smoking for non-participants $P(Y_i|M_{2i}=0)$ cannot be estimated using the observed data without making further assumptions. This may be written as
\begin{equation}
P(Y_i|M_{2i}=0) = \int\int P(Y_i|T_i,X_i,M_{2i}=0)P(T_i,X_i|M_{2i}=0) dX_i dT_i, \hspace{20pt} i \in \Omega \nonumber
\end{equation}
where $P(Y_i|T_i,X_i,M_{2i}=0)$ is not estimable but $P(T_i,X_i|M_{2i}=0)$ is estimable from observed data. We now assume that 
\begin{equation}
P(Y_i|T_i,X_i,M_{2i}=0) = P(Y_i|T_i,X_i,M_{2i}=1), \hspace{20pt} i \in \Omega \label{central_ass}
\end{equation}
which means that, given the observations $T_i$ and $X_i$, additional observation $M_{2i}=1$ or $M_{2i}=0$ does not give us any further understanding about the distribution of $Y_i$. Thus, for the rest of our paper, we restrict the models of interest to the cases for which the Equation \eqref{central_ass} holds. Now the smoking prevalence \eqref{marginal_distY} can be estimated if the probabilities $P(M_{2i} = 1)$, $P(Y_i|M_{2i} = 1)$, $P(M_{2i} = 0)$, $P(T_i,X_i|M_{2i}=0)$ and $P(Y_i|T_i,X_i,M_{2i}=1)$ can be estimated. The assumption \eqref{central_ass} can be justified if the relation $Y_i \rightarrow T_i$ is strong, i.e. the early onset of lung cancer or COPD is a strong indicator of smoking. In practice, the model parameters for relations of $X_i, Y_i$ and $T_i$ are estimated using data from the participants only.

\subsection{Construction of posterior distribution}

The model consists of two sub-models: a survival model for $T_i^*$, and a logistic regression model for the smoking indicator $Y_i^*$. Next, the parametric forms for sub-models are considered.

Time-to-disease variable $T_i^*|m_{1i}=1$ is assumed to follow Weibull distribution with a common shape parameter $a$ and scale parameter $b_i$ varying person by person. The distribution is left-truncated by the person's age $t_{0i} = x_{1i}$ at the beginning of follow-up. The likelihood contribution for observed disease cases can be written as 
$$p(T=t_{1i}|a,b,r_i=1,T>t_{0i}) = \frac{a b {t_{1i}}^{a-1} \text{exp}(-b {t_{1i}}^a)}{(1-F(t_{0i}))}\hspace{1cm}\text{for}\hspace{2mm}t_{1i}>t_{0i},$$ where $F(t)$ is cumulative distribution function for  Weibull distribution. For censored cases $i: r_i=0$ the likelihood contribution is the survival function $$S(T > t_{1i}|a,b,r_i=0,T>t_{0i}) = \text{exp}( -b({t_{1i}}^a-{t_{0i}}^a) )\hspace{1cm}\text{for}\hspace{2mm}t_{1i}>t_{0i}.$$

Parameter $b_i$ varies person by person based on the covariate measurements 
\begin{align}
\text{log}(b_i) = \gamma_0&+\gamma_1 x_{3i} + \gamma_2 Y_i + \gamma_3 x_{3i} Y_i \nonumber\\
&+ \gamma_{43}A_{3i} + \gamma_{44}A_{4i} + \gamma_{45}A_{5i} + \gamma_{46}A_{6i}\nonumber\\
&+ \gamma_{53}x_{3i}A_{3i} + \gamma_{54}x_{3i}A_{4i} + \gamma_{55}x_{3i}A_{5i} + \gamma_{56}x_{3i}A_{6i} \label{survival-regression} \\
&+ \gamma_{62} D_{2i} + \gamma_{63} D_{3i} + \gamma_{64} D_{4i} + \gamma_{65} D_{5i} + \gamma_{66} D_{6i} \nonumber\\ 
&+ \gamma_{72} x_{3i}D_{2i} + \gamma_{73} x_{3i}D_{3i} + \gamma_{74} x_{3i}D_{4i} + \gamma_{75} x_{3i}D_{5i} + \gamma_{76} x_{3i}D_{6i}, \nonumber
\end{align}
where parameter $\gamma_0$ corresponds to lung disease risk of non-smoking men at baseline (year 1972, North Karelia), $\gamma_1$ indicates the difference of risks for non-smoking men and women, $\gamma_2$ indicates the effect of smoking for men at baseline and $\gamma_3$ describes how disease risk for smoking women is different from the risk of smoking men (at baseline). The $\gamma_{42},\dots,\gamma_{46}$ stand for how the 
other areas differ from the baseline area (North Karelia) for men. 
The coefficients $\gamma_{53},\dots,\gamma_{56}$ describe how the last-mentioned quantities differ between the women and men. The $\gamma_{62},\dots,\gamma_{66}$ are the differences of the study year to the baseline study (year 1972) for men, and $\gamma_{72},\dots,\gamma_{76}$ are the differences of women and men for that particular study year. In Equation \eqref{survival-regression} the variables $A_{2i},\dots,A_{6i}$ are indicators for the study area such that $A_{2i}=1$ for the North Karelia (area $2$), $A_{3i}=1$ for the Northern Savonia (area $3$), $A_{4i}=1$ for Turku and Loimaa (area $4$), $A_{5i}=1$ for Helsinki and Vantaa (area $5$), and $A_{6i}=1$ for Oulu province (area $6$). Similarly, $D_{1i},\dots,D_{6i}$ are indicators about the study year such that $D_{1i}=1$ for 1972, $D_{2i}=1$ for 1977, $D_{3i}=1$ for 1982, $D_{4i}=1$ for 1987, $D_{5i}=1$ for 1992, and  $D_{6i}=1$ for 1997.

The smoking indicator is modelled also using logistic regression. The effects of gender $x_{3i}$, year of birth $x_{birth,i} = x_{4i}-x_{1i}$ and study year $x_{4i}$ are included in the model. We assume that the smoking indicator is Bernoulli distributed
\begin{equation}
Y_i \sim \text{Bernoulli}(s_i) \nonumber
\end{equation}
with probability $s_i$ such that
\begin{equation}
\text{logit}(s_i) = \alpha_{0,a,u,g} + \alpha_{1,a,u,g}(x_{birth,i}-1930),\label{logit_smoke} 
\end{equation}
where $a=x_{2i}$ is area, $g=x_{3i}$ is gender and $u=x_{4i}$ is study year for person $i$.
The coefficient $\alpha_{0,a,u,g}$ represents the intercept term for persons living in area $a$, of gender $g$, who were born in 1930 and were selected to the sample in year $u$. The year 1930 was chosen as a reference level because all the studies have some participants who were born in 1930. The coefficients $\alpha_{1,a,u,g}$ represents the impact of year of birth to the probability of smoking.

The information on the area (North Karelia or Northern Savonia) is missing for non-participants (2,664 in total) in 1972 and 1977. This missingness is due to accidentally lost data. These values are imported using multiple imputation with fixed probabilities $P(\text{area was Northern Savonia}|1972) = 0.495$ and $P(\text{area was Northern Savonia}|1977) = 0.493$. The imputation is not necessary for model fitting purposes, but is needed for the comparison of the areawise smoking trends.

\subsection{Model fitting and model diagnostics}
The model was built and fitted using JAGS \citep{jags-2003}, which is a tool for Bayesian analysis \citep{bayesian-data-analysis} of graphical models using Markov chain Monte Carlo (MCMC) \citep{robertcasella2004}. For all parameters the prior distributions were set as normal distributions with zero mean and variance $\sigma^2 = 1,000$. Regarding the scale of the parameters these priors are non-informative. Eight chains were run in parallel. Each of the chains had $200,000$ iterations from which the first $40,000$ were discarded as a burn-in. From the remaining $160,000$ iterations, the values of each $250$th iteration were stored to produce eight final thinned chains of $640$ iterations. In total we have $640*8 = 5,120$ realizations from posterior to use.

\begin{figure}[!htp]
\includegraphics[width=\textwidth]{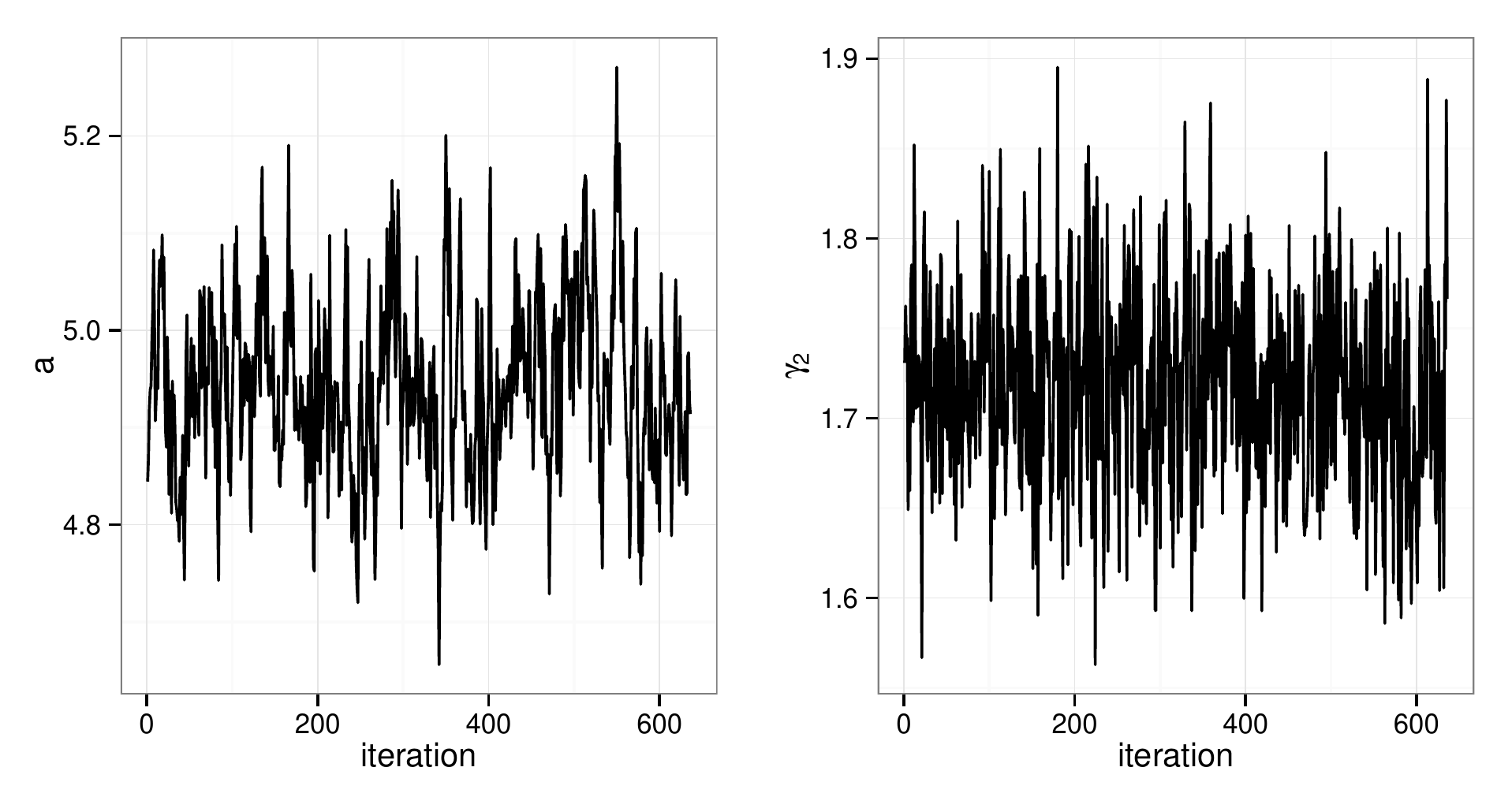}
\caption{Chain plots of MCMC computation. Left: Weibull shape-parameter $a$. Right: regression coefficient of smoking variable $\gamma_2$ of survival model.}
\label{final-mcmc}
\end{figure}

The MCMC convergence was monitored by Brooks-Gelman-Rubin convergence diagnostic \citep{brooks1998general}. The diagnostics of all parameters were below $1.01$ when values below $1.05$ indicate convergence. 
One of the MCMC chains of the final model is visualized for two parameters in the Figure \ref{final-mcmc}. The Figure shows that the Weibull shape-parameter is less well mixed than the other parameter. This is due to large autocorrelation caused by dependency on Weibull scale parameter. The better mixing on the smoking coefficient $\gamma_2$ is also visualized in the Figure. The majority of the parameters have good mixing. 
Posterior summaries of regression coefficients are given in Table \ref{posterior-table1} and Table \ref{posterior-table2}, see Appendix A.

The model diagnostics included a graphical comparison of the posterior predictive distribution against the observed values. The model was concluded to have a good fit to the data.

\section{Comparison of corrected and uncorrected smoking trends} 
\label{comparison-of-trends}

To obtain knowledge about the smoking prevalence for the study populations, we apply data-augmentation \citep{Tanner1987} to impute the missing values of smoking for non-participants, and take into account censoring of $T_i$. Because we apply Bayesian inference, the imputations are drawn from the posterior predictive distribution. First, the posterior samples of the regression coefficients are obtained using MCMC and participants data. Imputations of the smoking indicator for non-participants are drawn using the following procedure, which we implemented in R \citep{r}. The imputation depends on whether the event is observed or censored. If $T_i$ is censored, then first event-time $\tilde{T_i}$ for $T_i$ is generated using 
\begin{equation}
\tilde{T_i} \sim P(\tilde{T_i}|X_i) = P(\tilde{T_i}|X_i,Y_i^{'}=1)P(Y_i^{'}=1|X_i)+P(\tilde{T_i}|X_i,Y_i^{'}=0)P(Y_i^{'}=0|X_i).\nonumber
\end{equation} 
After that, use the imputed event-time $\tilde{T_i}$ to simulate $\tilde{Y_i} \sim P(\tilde{Y_i}|\tilde{T_i},X_i)$. If $T_i$ is observed, then simulate $\tilde{Y_i} \sim P(\tilde{Y_i}|T_i,X_i)$ straightforwardly based on the observed event-time.

\begin{figure}[!htp]
\includegraphics[width=\textwidth]{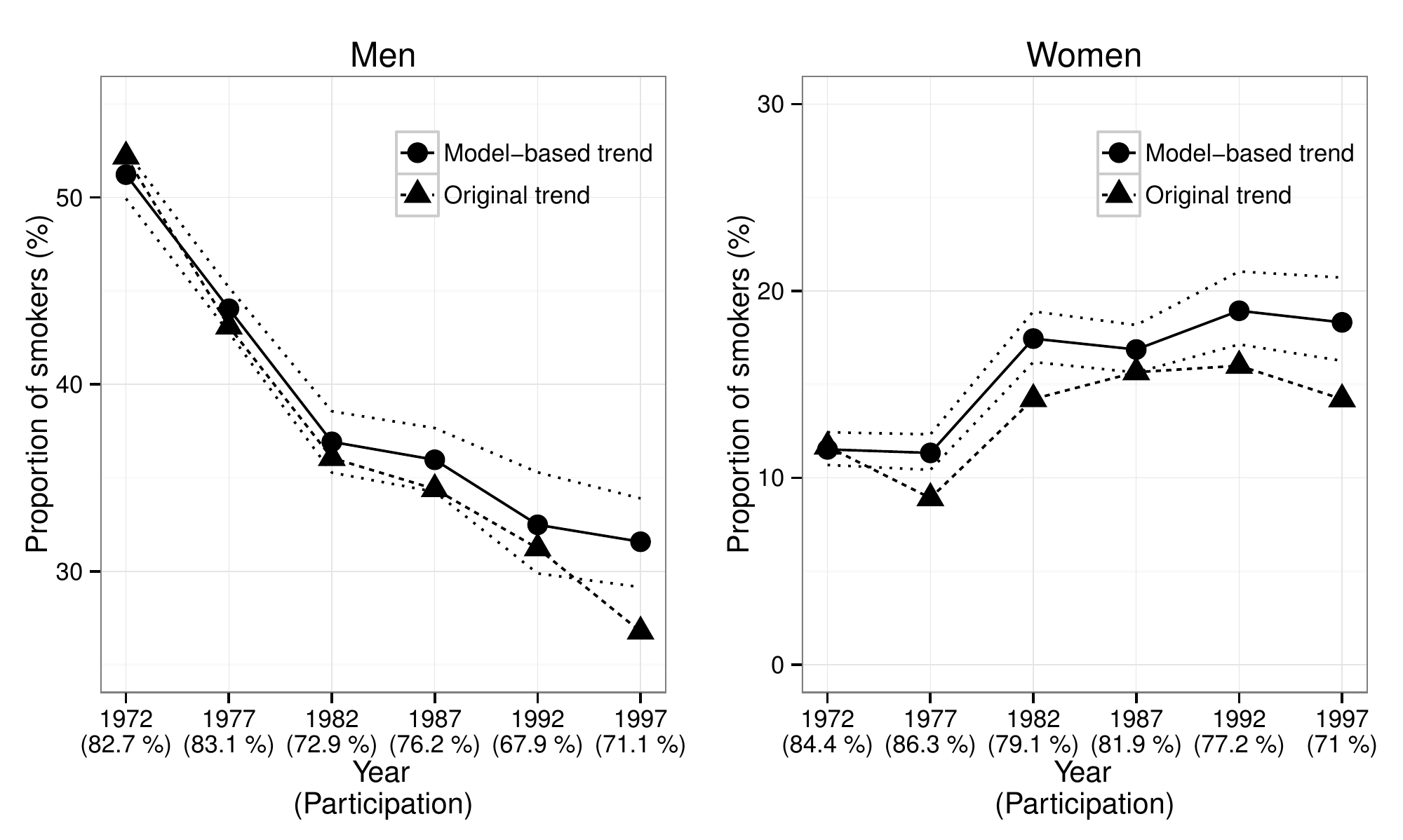}
\caption{Model-based trend and original trend for men (left) and women (right) in North Karelia province. The North Karelia was chosen because of the most visible change in the trends among the areas. Two dotted lines represent 95\% credible interval of the posterior distribution for corrected trends. Both the model-based and the original trend use WHO Scandinavian standardization weights.}
\label{model-based-trends}
\end{figure}

After the imputation, the survey sampling design has to be taken into account. We may treat data with each imputation as a full dataset. To provide area-specific population-level estimates we may then utilize inverse sampling probability weights \citep{lehtonen_pahkinen2004}. In addition to utilizing the sampling weights, the estimates were adjusted using WHO Scandinavian standardization weights \citep{ahmad2001age} in order to make the smoking rates internationally comparable. As an outcome, we obtain area-specific trend estimates for both genders corresponding to each imputation. These trends can be considered as samples from the posterior distribution of the trends. The estimated model-based corrected trends are compared to the corresponding original trends in Figure \ref{model-based-trends} for North Karelia. The original or uncorrected trends were produced from the participant data only. The adjustment for sampling design and the WHO weights was the same as for the model-based trends.

In Figure \ref{model-based-trends} the difference between the trends increases as the participation rate decreases. In addition, it seems that the difference of the trends in most time-points is larger for women than for men. On the other hand, the largest difference in the corrected and non-corrected prevalence estimates is $6.6$ percentage points (relative difference of $25\%$), which is observed for men in Helsinki and Vantaa in 1997. The comparison of the model-based and original smoking prevalence trends for the study year 1997 is presented in Table \ref{comparison-1997}.

\begin{table}[!htbp] \centering 
  \caption{Observed and model-based smoking proportions for the study in 1997 adjusted using WHO Scandinavian standardization weights. The two rightmost columns describe the 95\% credible intervals of model-based trends. Participant smoking is the same as "Original trend" in Figure \ref{model-based-trends}} 
  \label{comparison-1997} 
\begin{tabular}{@{\extracolsep{0pt}} llccrr} 
\\[-1.8ex]\hline 
\hline \\[-1.8ex] 
Gender & Area &  \shortstack{Participant\\ smoking (\%)} & \shortstack{Model-based\\ total smoking (\%)} &  \multicolumn{2}{l}{95\% Credible Interval} \\\hline 
  Men &        North Karelia & 26.8 & 31.6 & 29.2 & 33.9 \\
  Men &     Northern Savonia & 30.7 & 31.8 & 29.5 & 34.0 \\
  Men &     Turku and Loimaa & 32.4 & 33.7 & 31.1 & 36.2 \\
  Men &  Helsinki and Vantaa & 26.1 & 32.7 & 30.1 & 35.6 \\
  Men &        Oulu province & 30.1 & 32.3 & 29.5 & 35.2 \\
Women &        North Karelia & 14.2 & 18.3 & 16.3 & 20.7 \\
Women &     Northern Savonia & 17.0 & 19.1 & 17.1 & 21.1 \\
Women &     Turku and Loimaa & 20.5 & 23.6 & 21.3 & 26.0 \\
Women &  Helsinki and Vantaa & 22.6 & 27.7 & 25.3 & 30.4 \\
Women &        Oulu province & 19.1 & 22.2 & 20.0 & 24.3\\
\hline \\[-1.8ex] 
\end{tabular}
\end{table}

\section{Discussion} 
\label{discussion}

We have proposed an approach to overcome the challenges with non-ignorable missing data in epidemiological studies and have applied it to estimate the population trends of smoking in Finland in 1972--1997. The approach uses follow-up data to obtain information on risk factors missing at baseline. Thanks to the administrative registers in Finland, the follow-up data are available also for non-participants. Smoking has been selected as the risk factor of interest because it is a strong risk factor of lung cancer and COPD and potentially has an effect to the decision on the HES participation.

We evaluated the proportion of smokers combining the available information from both the participants and non-participants for the FINRISK study. Our results indicate that the levels of smoking prevalence is affected when the information provided by lung cancer and COPD time-to-event data is accounted to provide an estimate for the smoking of non-participants.

In general, statistical modelling under the non-ignorable missingness requires external information on the missingness mechanism. It can be argued that the inclusion of follow-up data provides the information needed. The situation can be formally described using causal models with design and then modelled by a Bayesian model. The idea of utilizing existing causal knowledge to fix non-ignorable missingness is not restricted to survival models. 

The approach is limited by the availability of follow-up data. It takes years or decades until the follow-up data on lung cancer and COPD can be used to model the missing data mechanism. It is unclear to what extent the approach can be applied in other countries because register-based baseline and follow-up data sets are not usually available for non-participants. Although the approach may not be directly applicable in a study, the results from other similar studies, where the approach has been applied, may provide a starting point for the prior setting and the sensitivity analyses.

Censoring was treated as non-informative, which may cause some bias to the estimates. As smoking is a risk factor for many common causes of death, an individual censored due to other deaths is more likely to be a smoker than an individual censored due to the end of the follow-up. It is therefore expected that the actual proportions of smokers could be even higher than the corrected proportions reported here. Improved estimation would require a competing risk approach with a a comprehensive set of risk factors and a number of disease-specific endpoints. This is left as future work.

Inclusion of information about smoking as a time-dependent process would yield more realistic expressions of smoking in different age-groups. The effect of smoking years could be then considered as a covariate for the lung diseases. With the current model, it is assumed that observed lung disease diagnosis e.g. at age 50 is equally strong indication about smoking, no matter if the person is diagnosed five or 25 years after the survey. In reality, individuals may have started or stopped smoking after the survey was conducted.

The presented approach may be utilized with data arising in forthcoming FINRISK surveys. In addition, the model could be used to give recommendations on the sample size and the stratification. 

Our work reminds that data with MNAR-situation may be changed to MAR using additional assumption and external information. This allows us to provide estimates that describe the whole population instead of the restricted sample of survey participants.

\section*{Acknowledgements}
The work was supported by the Academy of Finland [grant number 266251].

\bibliographystyle{wb_stat}
\bibliography{viitteet3}

\newpage
\section*{Appendix A: Regression coefficients}
\begin{table}[!htbp] \centering 
\small
  \caption{Posterior summaries of the estimated parameters for the smoking model, reduced to the parameters of North Karelia.}
  \label{posterior-table1} 
\begin{tabular}{@{\extracolsep{-5pt}} lcrrrr} 
\hline 
\hline \\[-1.8ex] 
Description of related variable & Parameter & Mean & SD & 2.5\% & 97.5\% \\\hline
men born at 1930 in 1972 & $\alpha_{0,1972,1,2}$ & $0.086$ & $0.043$ & $0.003$ & $0.168$ \\ 
men born at 1930 in 1977 & $\alpha_{0,1977,1,2}$ & $-0.294$ & $0.047$ & $-0.384$ & $-0.203$ \\ 
men born at 1930 in 1982 & $\alpha_{0,1982,1,2}$ & $-0.672$ & $0.069$ & $-0.806$ & $-0.538$ \\ 
men born at 1930 in 1987 & $\alpha_{0,1987,1,2}$ & $-0.888$ & $0.084$ & $-1.052$ & $-0.724$ \\ 
men born at 1930 in 1992 & $\alpha_{0,1992,1,2}$ & $-1.092$ & $0.159$ & $-1.409$ & $-0.779$ \\ 
men born at 1930 in 1997 & $\alpha_{0,1997,1,2}$ & $-1.449$ & $0.107$ & $-1.661$ & $-1.244$ \\ 
women born at 1930 in 1972 & $\alpha_{0,1972,2,2}$ & $-2.106$ & $0.070$ & $-2.242$ & $-1.971$ \\ 
women born at 1930 in 1977 & $\alpha_{0,1977,2,2}$ & $-2.452$ & $0.086$ & $-2.625$ & $-2.287$ \\ 
women born at 1930 in 1982 & $\alpha_{0,1982,2,2}$ & $-2.361$ & $0.111$ & $-2.583$ & $-2.153$ \\ 
women born at 1930 in 1987 & $\alpha_{0,1987,2,2}$ & $-2.412$ & $0.128$ & $-2.670$ & $-2.164$ \\ 
women born at 1930 in 1992 & $\alpha_{0,1992,2,2}$ & $-2.599$ & $0.219$ & $-3.039$ & $-2.183$ \\ 
women born at 1930 in 1997 & $\alpha_{0,1997,2,2}$ & $-2.833$ & $0.204$ & $-3.239$ & $-2.449$ \\ 
difference of year of birth to 1930 (men in 1972) & $\alpha_{1,1972,1,2}$ & $0.001$ & $0.004$ & $-0.007$ & $0.009$ \\ 
difference of year of birth to 1930 (men in 1977) & $\alpha_{1,1977,1,2}$ & $0.008$ & $0.005$ & $-0.0004$ & $0.017$ \\ 
difference of year of birth to 1930 (men in 1982) & $\alpha_{1,1982,1,2}$ & $0.013$ & $0.005$ & $0.003$ & $0.022$ \\ 
difference of year of birth to 1930 (men in 1987) & $\alpha_{1,1987,1,2}$ & $0.019$ & $0.005$ & $0.009$ & $0.028$ \\ 
difference of year of birth to 1930 (men in 1992) & $\alpha_{1,1992,1,2}$ & $0.017$ & $0.007$ & $0.002$ & $0.032$ \\ 
difference of year of birth to 1930 (men in 1997) & $\alpha_{1,1997,1,2}$ & $0.029$ & $0.005$ & $0.019$ & $0.038$ \\ 
difference of year of birth to 1930 (women in 1972) & $\alpha_{1,1972,2,2}$ & $0.043$ & $0.007$ & $0.030$ & $0.056$ \\ 
difference of year of birth to 1930 (women in 1977) & $\alpha_{1,1977,2,2}$ & $0.050$ & $0.008$ & $0.034$ & $0.066$ \\ 
difference of year of birth to 1930 (women in 1982) & $\alpha_{1,1982,2,2}$ & $0.057$ & $0.007$ & $0.044$ & $0.070$ \\ 
difference of year of birth to 1930 (women in 1987) & $\alpha_{1,1987,2,2}$ & $0.049$ & $0.006$ & $0.037$ & $0.062$ \\ 
difference of year of birth to 1930 (women in 1992) & $\alpha_{1,1992,2,2}$ & $0.049$ & $0.009$ & $0.031$ & $0.066$ \\ 
difference of year of birth to 1930 (women in 1997) & $\alpha_{1,1997,2,2}$ & $0.049$ & $0.007$ & $0.035$ & $0.064$ \\ 
\hline \\[-1.8ex]
\end{tabular} 
\end{table}

\begin{table}[!htbp] \centering 
\small
  \caption{Posterior summaries of the estimated parameters for the survival model (includes all parameters).}
  \label{posterior-table2} 
\begin{tabular}{@{\extracolsep{-5pt}} lcrrrr} 
\hline 
\hline \\[-1.8ex]
Description of related variable & Parameter & Mean & SD & 2.5\% & 97.5\% \\\hline
Weibull shape-parameter & $a$ & $4.257$ & $0.111$ & $4.041$ & $4.475$ \\ 
intercept (men) & $\gamma_0$ & $-21.848$ & $0.501$ & $-22.817$ & $-20.859$ \\ 
gender (women) & $\gamma_1$ & $-1.352$ & $0.164$ & $-1.665$ & $-1.031$ \\ 
smoking & $\gamma_2$ & $1.772$ & $0.061$ & $1.653$ & $1.893$ \\ 
interaction of smoking and gender & $\gamma_3$ & $0.559$ & $0.116$ & $0.328$ & $0.786$ \\ 
Northern Savonia & $\gamma_{43}$ & $0.070$ & $0.054$ & $-0.036$ & $0.175$ \\ 
Turku and Loimaa & $\gamma_{44}$ & $-0.298$ & $0.085$ & $-0.466$ & $-0.134$ \\ 
Helsinki and Vantaa & $\gamma_{45}$ & $-0.389$ & $0.131$ & $-0.652$ & $-0.139$ \\ 
Oulu province & $\gamma_{46}$ & $-1.290$ & $0.300$ & $-1.931$ & $-0.743$ \\ 
interaction of Northern Savonia and women & $\gamma_{53}$ & $-0.274$ & $0.131$ & $-0.528$ & $-0.023$ \\ 
interaction of Turku and Loimaa and women & $\gamma_{54}$ & $0.654$ & $0.161$ & $0.337$ & $0.970$ \\ 
interaction of Helsinki and Vantaa and women & $\gamma_{55}$ & $0.509$ & $0.241$ & $0.037$ & $0.963$ \\ 
interaction of Oulu province and women & $\gamma_{56}$ & $1.072$ & $0.477$ & $0.104$ & $2.006$ \\ 
year 1977 & $\gamma_{62}$ & $-0.242$ & $0.074$ & $-0.382$ & $-0.094$ \\ 
year 1982 & $\gamma_{63}$ & $0.017$ & $0.074$ & $-0.125$ & $0.164$ \\ 
year 1987 & $\gamma_{64}$ & $-0.090$ & $0.105$ & $-0.295$ & $0.117$ \\ 
year 1992 & $\gamma_{65}$ & $-0.185$ & $0.126$ & $-0.433$ & $0.062$ \\ 
year 1997 & $\gamma_{66}$ & $0.134$ & $0.107$ & $-0.075$ & $0.345$ \\ 
interaction of women and year 1977 & $\gamma_{72}$ & $0.269$ & $0.162$ & $-0.042$ & $0.582$ \\ 
interaction of women and year 1982 & $\gamma_{73}$ & $-0.240$ & $0.174$ & $-0.581$ & $0.092$ \\ 
interaction of women and year 1987 & $\gamma_{74}$ & $0.182$ & $0.212$ & $-0.234$ & $0.600$ \\ 
interaction of women and year 1992 & $\gamma_{75}$ & $0.506$ & $0.225$ & $0.058$ & $0.950$ \\ 
interaction of women and year 1997 & $\gamma_{76}$ & $0.343$ & $0.212$ & $-0.078$ & $0.753$ \\ 
\hline \\[-1.8ex] 
\end{tabular} 
\end{table}

\end{document}